\documentclass[12pt,english]{aastex}
\usepackage[T1]{fontenc}
\usepackage[latin9]{inputenc}
\usepackage{amssymb}
\usepackage{natbib}
\makeatletter

\providecommand{\tabularnewline}{\\}

\makeatother

\usepackage{babel}
\usepackage{array}

\begin{document}

\title{A Domain Wall Solution by Perturbation of the Kasner Spacetime}

\author{George Kotsopoulos}

\affil{1906-373 Front St. West, Toronto, ON M5V 3R7}

\email{g.kotsopoulos@utoronto.ca}

\and

\author{Charles C. Dyer}

\affil{Dept. of Physical and Environmental Sciences, University of
Toronto Scarborough, and Dept. of Astronomy and Astrophysics, University
of Toronto, Toronto, Canada}

\email{dyer@astro.utoronto.ca}

\date{January 24, 2012}

\begin{abstract}
Plane symmetric perturbations are applied to an axially symmetric Kasner
spacetime which leads to no momentum flow orthogonal to the planes of
symmetry. This flow appears laminar and the structure can be interpreted
as a domain wall.  We further extend consideration to the class of
Bianchi Type I spacetimes and obtain corresponding results.
\end{abstract}


\section{Finding New Solutions}

There are three principal exact solutions to the Einstein Field Equations (EFE) that are most relevant for the description of astrophysical
phenomena. They are the Schwarzschild (internal and external), Kerr and Friedman-LeMa\^{i}tre-Robertson-Walker (FLRW) models. These solutions are all highly idealized and involve the introduction of simplifying assumptions such as symmetry.

The technique we use which has led to a solution of the EFE involves the
perturbation of an already symmetric solution~\citep{Wilson}. Whereas
standard perturbation techniques involve specifying the forms of the energy-momentum tensor to determine the form of the metric, we begin by
defining a new metric with an applied perturbation:
\begin{equation}
g_{ab}=\tilde{g}_{ab}+h_{ab}\end{equation}
\noindent where $\tilde{g}_{ab}$ is a known symmetric metric and
$h_{ab}$ is the applied perturbation. We then use this new metric
to find the components of the Einstein tensor, $G_{ab}$, in terms
of the metric components. Next, we invert the EFE, where now $T_{ab}=G_{ab}/\kappa$,
to ascertain whether or not components of the energy-momentum tensor,
$T_{ab}$, exist to satisfy the perturbation. If such an energy-momentum
tensor is physically acceptable, we then try to limit the behaviour of our perturbation
by imposing conditions to model particular astronomical phenomenon.
For example, if we were modelling a galaxy-like structure, we could
choose our density to scale as $\rho \sim r^{-2}$ from the
galactic centre~\citep{Wilson}. This makes the technique flexible and relevant in that
we can model a physically viable structure. With these real physical
constraints applied, we have a new metric. We now
re-compute the energy-momentum components of this new metric and investigate
any implications that may arise from our new exact
solution to the EFE.
\section{The Base Solution\label{KASNER}}

The metric we choose to perturb is the Kasner spacetime~\citep{Kasner}. This metric describes an anisotropic universe that is a vacuum solution to the EFE classified  as a Bianchi Type I universe. It is characterized
by a spacetime that is anisotropically expanding (or contracting)
in two directions while contracting (or expanding) in the third with
space-like slices that are spatially flat and with a singularity occurring
at $t=0$. Hence, the Kasner spacetime is of special interest in cosmology
since the standard cosmological solutions to the EFE near the cosmological
singularity, such as the aforementioned FLRW model, can be described
as a succession of Kasner epochs~\citep{Lifshitz}. Furthermore, by breaking
a homogeneity along one direction, Bianchi Type solutions can easily
be generalized to spacetimes with $G_{2}$ isotropy groups~\citep{Harvey}.

The Kasner spacetime in Cartesian coordinates with metric signature $(+,-,...,-)$
takes the form (in spacetime dimension $D$):
\begin{equation}
ds^{2}=dt^{2}-\sum_{j=1}^{D-1}t^{2P_{j}}[dx_{j}]^{2}\end{equation}
\noindent where $P_{j}$ are the Kasner exponents which satisfy:
\begin{equation}
\sum_{j=1}^{D-1}P_{j}=1
\;\;\;\;
\textrm{and}
\;\;\;\;
\sum_{j=1}^{D-1}P_{j}^{2}=1
\end{equation}
The first condition in (3) describes a plane whereas the
second condition describes a sphere of dimension $D-1$. Thus, the Kasner exponents lie on a sphere of dimension $D-2$. For $D=4$, at each $t=constant$ hypersurface, there exists a flat
3-dimensional space, whose worldlines of constant $x$,
$y$ and $z$ are time-like geodesics along which galaxies, or other
test particles, can be imagined to move. Since the solution represents
an anisotropically expanding (or contracting) universe, a volume element
$dV$ increases (or decreases) in time as $\sqrt{|g|}d^{3}x=td^{3}x$,
where $g=det|g_{ab}|$.

The Kasner spacetime in Cartesian coordinates in a $4-D$ spacetime takes
the form:
\begin{equation}
ds^{2}=dt^{2}-t^{2P_{1}}dx^{2}-t^{2P_{2}}dy^{2}-t^{2P_{3}}dz^{2}\end{equation}
\noindent with the same restrictions imposed by the Kasner exponents
as given in (3).

The form of the Kasner spacetime that we consider is the
axisymmetric case with $G_{2}$ isometry that occurs when two of
the exponents are equal, which by (3), requires that $P_{1}=P_{2}=2/3$ and
$P_{3}=-1/3$. With this restriction, the metric in cylindrical coordinates becomes:
\begin{equation}
ds^{2}=dt^{2}-t^{4/3}(dr^{2}+r^{2}d\phi^{2})-t^{-2/3}dz^{2}\end{equation}

\section{The Perturbations}

The set of perturbations we implement are only on the $z$-component of
the Kasner spacetime represented in cylindrical coordinates (5). Clearly,
the results we obtain in cylindrical coordinates will be similar to
those in Cartesian coordinates since $r^{2}=x^{2}+y^{2}$. For simplicity, we define the Kasner spacetime in cylindrical coordinates as:
\begin{equation}
ds^{2}=dt^{2}-t^{4/3}(dr^{2}+r^{2}d\phi^{2})-g_{zz}dz^{2}\end{equation}
\noindent where $g_{zz}=\tilde{g}_{zz}+h_{ab}$ is our perturbed metric. We consider the three Cases shown in Table 1.
\begin{center}
\begin{tabular}{|c|c|}
\hline 
Cases & $g_{zz}=\tilde{g}_{zz}+h_{ab}$\tabularnewline
\hline
\hline 
I & $[t^{-2/3}+H]$\tabularnewline
\hline 
II & $[t^{-1/3}+H]^{2}$\tabularnewline
\hline 
III & $[t^{-2/3}+Z(z)T(t)]$\tabularnewline
\hline
\end{tabular}
\par\end{center}
\begin{center}
Table 1: Implemented perturbations.
\par\end{center}
\noindent
In Cases I and II, we first investigate $H$ as a differentiable function of $z$. We then investigate Cases I and II again,
but where $H$ is now made a differentiable function of $t$, $r$ and $z$. In Case III,
we investigate the result of a perturbation with a product separable differentiable solution.

To aid in these computations, we use the REDUCE Computer Algebra
System ~\citep{Hearn} with the REDTEN Tensor Analysis Package ~\citep{Harper}.

\section{Results of the Perturbations}

Both Cases I and II produce a $G_{ab}$
with non-zero components. When the perturbation is $H(z)$,
the result produces non-zero terms for $G_{tt}$, $G_{rr}$ and $G{}_{\phi\phi}$.
All other terms, including $G_{zz}$, are zero. When the perturbation
is $H(t,r,z)$, a similar result is obtained for all cases, but with the
addition of cross-components, $G_{ab}$ for $a\neq b$. These cross
terms can be made zero if we choose the perturbation to contain only
first-order terms in $H$ or by making $\partial H/\partial r=\partial H/\partial\phi=0$.
Also, in cylindrical coordinates for all the cases reviewed,
$G_{rr}=r^{2}G_{\phi\phi}$, which is expected. The product
separable perturbation of Case III reveals the same results as Case
I and II; we arrive at non-zero terms for $G_{tt}$, $G_{rr}$
and $G{}_{\phi\phi}$ and a $G_{zz}=0$. Thus all perturbations
that involved only the $g_{zz}$ term of the Kasner metric
reveal that we will always obtain $G_{zz}=0$ when considering
axisymmetric or plane symmetric Kasner spacetimes, when $P1=P2$.

Relating these results to the energy-momentum tensor $T_{ab}$, both
the $G_{zz}$ and $G_{za}=G_{az}$ (for $a=\{0,1,2\}$) components
of the Einstein tensor being zero reveals that there is \emph{no}
momentum-flux across the $z=constant$ surface regardless of the type
of perturbation applied. Each
of the applied perturbations resulted in the metric collapsing (or expanding) in
the $z$-direction while expanding (or collapsing) in the $x$- and $y$-directions,
as is expected of the Kasner metric. But, as a result of the $G_{zz}$
term being zero, there is no interaction between the stratified layers
of the matter above or below the $xy$-plane. This is suggestive of laminar flow.

Inverting the EFE to get $T_{ab}=G_{ab}/\kappa$, leads to $T_{ab}$ being diagonal but with no $T_{zz}$ component.
The energy-momentum tensor for an infinite, static plane-symmetric domain wall as suggested by Campanelli et al is:
\begin{equation}
T_{ab}=\delta(z)diag(\rho,\,-p,\,-p,\,0)\end{equation}
where $\rho$ is the energy-density and $p$ is the pressure. This
energy-momentum tensor~\citep{Campanelli} corresponds to an infinite, static plane-symmetric
domain wall~\citep{Kibble} lying in the $xy$-plane. Domain walls correspond
to a particular class of topological defects whereby the energy-density
is trapped, and from a cosmological point of view, these over-dense
regions could lead to structure formation~\citep{Brandenberger}. Furthermore, it has been suggested~\citep{Friedland} that the Universe may be dominated by a network of domain walls and these domain wall structures could represent an alternative view of dark energy theories.

\section{Energy Conditions}

In order to rule out any non-physical solutions to the EFE, energy
conditions are applied to the state of matter content for gravitational
and non-gravitational fields. These energy conditions consist of the
Weak, Null, Strong and Dominant energy conditions, which are
coordinate-invariant constraints on the energy-momentum tensor.

For the cases in which our Kasner spacetime was perturbed with $H(z)$,
we apply the least stringent of these conditions, the Null Energy
Condition (NEC). This condition states that for all future-pointing
null vectors, $k^{a}$:
\begin{equation}
T_{ab}k^{a}k^{b}\geqq0\end{equation}
\noindent This restriction implies that $\rho+p\geqq0$, whereby the
energy-density may be negative as long as there is a compensating
pressure. If this condition is violated in any of the perturbed Kasner
spacetimes, it would then indicate that our solutions are unstable.

We choose the null vector field $k^{a}$, in the Kasner spacetime to be,
\begin{equation}
k^{a}=\left(\sqrt{\frac{1-n^{2}g_{zz}}{g_{tt}}},0,0,n^{2}\right)\end{equation}
\noindent where $n\,\mathbb{\epsilon\, R}$. This class of null vectors lies in the $T_{zz}$ plane of interest. A summary of the energy-density
restrictions is shown in Table 2, where $\dot{T}=\partial T/\partial t$. For all Cases I to III, $T_{ab}k^{a}k^{b}$ is positive definite for positive $H$.

\noindent \begin{center}
\setlength{\extrarowheight}{10pt}
\setlength{\tabcolsep}{15pt}
\begin{tabular}{|c|c|c|}
\hline
Cases & \large$g_{zz}=\tilde{g}_{zz}+h_{ab}$ & \large$T_{ab}k^{a}k^{b}$\tabularnewline
\hline
\hline
I & $[t^{-2/3}+H(z)]$ & \Large$\frac{4t^{1/3}(t^{1/3}n^{2}+t)H}{9t^{4/3}(6t^{2/3}H+1)}$\tabularnewline
\hline 
II & $[t^{-1/3}+H(z)]^{2}$ & \Large$\frac{4(t^{1/3}n^{2}+t)H}{9t^{4/3}(t^{1/3}H+1)}$\tabularnewline
\hline 
III & $t^{-2/3}+H(z)T(t)$ &
\Large$\frac{2t^{1/3}(3t^{1/3}\dot{T}n^{2}t+3\dot{T}t^{2}+2t^{1/3}Tn^{2}+2Tt)H}{9t^{4/3}(6t^{2/3}HT+1)}$\tabularnewline
\hline
\end{tabular}
\par\end{center}
\begin{center}
Table 2: Applied NEC results
\par\end{center}

\section{Bianchi Type I Metrics}

As mentioned in Section \ref{KASNER}, the Kasner spacetime is a special class of Bianchi Type I spacetime being a homogeneous and anisotropic vacuum
solution to the EFE. Therefore, it is appropriate to apply the perturbations
to the more generalized Bianchi Type I to investigate its dynamics
and compare the results to the perturbed Kasner models. The Bianchi
Type I~\citep{Stephani} metric has the form:
\begin{equation}
\label{BianchiI}
ds^{2}=c^{2}dt^{2}-g_{11}dx^{2}-g_{22}dy^{2}-g_{33}dz^{2}\end{equation}
\noindent
where,
\begin{equation}
g_{\alpha\alpha}=(-g)^{1/3}[ct/(\hat{M}ct+A)]^{2P_{\alpha}-2/3}\end{equation}
such that $\alpha=1, 2,3$, with no sum on $\alpha$. The quantity $\hat{M}=\kappa\mu c^{2}\sqrt{-g}$ is a constant and $\sqrt{-g}=3ct(\hat{M}ct+A)/4,$
where $A$ is an integration constant. Using these relationships and an appropriate constant rescaling of coordinates, we can re-write (\ref{BianchiI}) as:
\begin{equation}
ds^{2}=dt^{2}-t^{4/3}dx^{2}-t^{4/3}dy^{2}-t^{-2/3}(1+\epsilon t)^{2}dz^{2}\end{equation}
\noindent where $\epsilon=\hat{M}/A$. As before, since we wish to compare this
result to the axisymmetric perturbed Kasner spacetime, we impose
$P_{1}=P_{2}=2/3$ and $P_{3}=-1/3$.

There are two limiting cases of the Bianchi Type I metric in (12).
If we choose $\epsilon t\ll1$, the initial singularity is approached
at early times and we regain the original Kasner spacetime. For late times,
if we choose $\epsilon t\gg1$, the metric isotropizes to the well-known
Einstein-de Sitter dust metric, which is a sub-set of the FLRW metric.
Indeed, the Kasner solution is a past asymptotic state as mentioned in Section \ref{KASNER}.

\section{Perturbations on Bianchi Type I}

The perturbations applied to the Bianchi Type I metric are summarized in Table 3.

\newpage
\noindent \begin{center}
\begin{tabular}{|c|c|}
\hline 
Cases & $g_{zz}=\tilde{g}_{zz}+h_{ab}$\tabularnewline
\hline
\hline 
I & $t^{-2/3}[(1+\epsilon t)^{2}+H(z)]$\tabularnewline
\hline 
II & $t^{-2/3}[(1+\epsilon t)^{2}+H(z)T(t)]$\tabularnewline
\hline
\end{tabular}
\par\end{center}
\begin{center}
Table 3: Implemented Perturbations on Bianchi Type I
\par\end{center}
\noindent
Investigating $G_{ab}$ for the perturbations
of Cases I and II, we arrive at the same conclusion as was
found for our perturbed Kasner spacetimes; we obtain
$G_{zz}=0$ for the axisymmetric Bianchi Type I spacetimes.

\section{Symmetries}

The nonlinear nature of the EFE makes it difficult to find exact solutions.
All of the known solutions have admitted simplifying symmetries in
order to attain a solution. The technique we apply involves the perturbation
of an already highly symmetric spacetime in hopes of producing a new spacetime
solution. We now wish to determine to what extent any symmetries that were inherent in the original spacetime
metric still retain (or break) any symmetries.

To investigate symmetries, we will consider conformal Killing vector
fields, $\xi^{a}$, that satisfy the conformal Killing equations:
\begin{equation}
\xi_{a||b}+\xi_{b||a}=\phi g_{ab}\end{equation}
where $\phi=\frac{1}{2}\xi_{\,\,||c}^{c}$ and $||$ denotes covariant differentiation.
Utilizing the REDUCE/REDTEN computer algebra system, we computed the
conformal Killing equations as 10 symmetric rank-2 tensors for each of the
perturbation cases investigated. A Killing vector from the original metric was then
calculated along with its covariant derivative and substituted in the conformal
Killing equation (13). For the perturbations applied to the two Cases, no such Killing vectors or conformal Killing vectors were found to remain; the original Killing vector is no longer a Killing vector.

\section{Conclusion}

Starting from two known, highly symmetric, solutions to the Einstein
Field Equations we have applied plane symmetric perturbations and have shown
that the resulting perturbed spacetimes exhibit the existence of
structure that can be interpreted as a domain wall. It was demonstrated
that these solutions do not violate the Null Energy Condition and thus
would permit us to apply an appropriate energy-momentum tensor in future
investigations.

\section{Acknowledgements}

This research was supported in part by the Natural Sciences and Engineering
Research Council of Canada via a Discovery Grant to CCD.

\bibliography{KasnerDomainWall}

\begin{thebibliography}{11}
\providecommand{\natexlab}[1]{#1}
\providecommand{\url}[1]{\texttt{#1}}
\expandafter\ifx\csname urlstyle\endcsname\relax
  \providecommand{\doi}[1]{doi: #1}\else
  \providecommand{\doi}{doi: \begingroup \urlstyle{rm}\Url}\fi

\bibitem[Brandenberger(1997)]{Brandenberger}
R.~Brandenberger.
\newblock Topological defects and the formation of structure in the universe.
\newblock \emph{astro-ph}, 211:\penalty0 105--+, 1997.

\bibitem[Campanelli et~al.(2003)Campanelli, Cea, Fogli, and
  Tedesco]{Campanelli}
L.~Campanelli, P.~Cea, G.~L. Fogli, and L.~Tedesco.
\newblock Gravitational field of static thin planar walls in weak-field
  approximation.
\newblock \emph{International Journal of Modern Physics D}, 12:\penalty0 1385,
  2003.
\newblock URL \url{doi:10.1142/S0218271803003773}.

\bibitem[Friedland et~al.(2003)Friedland, Murayama, and Perelstein]{Friedland}
A.~Friedland, H.~Murayama, and M.~Perelstein.
\newblock Domain walls as dark energy.
\newblock \emph{\prd}, 67\penalty0 (4):\penalty0 043519--+, February 2003.
\newblock \doi{10.1103/PhysRevD.67.043519}.

\bibitem[Harper and Dyer(1994)]{Harper}
John Harper and Charles~C. Dyer.
\newblock Redten, 1994.
\newblock URL \url{http://www.utsc.utoronto.ca/~harper/redten.html}.

\bibitem[Harvey(1990)]{Harvey}
Alex Harvey.
\newblock Will the real kasner metric please stand up.
\newblock \emph{General Relativity and Gravitation}, 22:\penalty0 1433--1445,
  1990.
\newblock ISSN 0001-7701.
\newblock URL \url{http://dx.doi.org/10.1007/BF00756841}.
\newblock 10.1007/BF00756841.

\bibitem[Hearn(2009)]{Hearn}
Anthony~C. Hearn.
\newblock Reduce, 2009.
\newblock URL \url{http://www.reduce-algebra.com/}.

\bibitem[Kasner(1921)]{Kasner}
Edward Kasner.
\newblock Geometrical theorems on einstein's cosmological equations.
\newblock \emph{American Journal of Math}, 43:\penalty0 217--221, 1921.

\bibitem[Kibble(1976)]{Kibble}
T.~W.~B. Kibble.
\newblock Topology of cosmic domains and strings.
\newblock \emph{Journal of Physics A Mathematical General}, 9:\penalty0
  1387--1398, August 1976.
\newblock \doi{10.1088/0305-4470/9/8/029}.

\bibitem[Lifshitz and Khalatnikov(1963)]{Lifshitz}
E.~M. Lifshitz and I.~M. Khalatnikov.
\newblock Investigations in relativistic cosmology.
\newblock \emph{Advances in Physics}, 12:\penalty0 185--249, April 1963.
\newblock \doi{10.1080/00018736300101283}.

\bibitem[Stephani(2003)]{Stephani}
Hans Stephani.
\newblock \emph{General Relativity: An Introduction to Special and General
  Relativity}.
\newblock Cambridge University Press, 2003.

\bibitem[Wilson and Dyer(2007)]{Wilson}
Brian Wilson and Charles~C. Dyer.
\newblock A galaxy-like perturbation of the robertson-walker metric.
\newblock \emph{General Relativity and Gravitation}, 39:\penalty0 2001--2015,
  December 2007.
\newblock \doi{10.1007/s10714-007-0497-0}.

\end{thebibliography}

\end{document}